# PyZoBot: A Platform for Conversational Information Extraction and Synthesis from Curated Zotero Reference Libraries through Advanced Retrieval-Augmented Generation.

[1,2]Suad Alshammari, Pharm.D., [2,3]Walaa Abu Rukbah, Pharm.D., [2,4]Lama Basalelah, Pharm.D., [2,5]Ali Alsuhibani, Pharm.D., [2,6]Ali Alghubayshi, Pharm.D., [7] Bridget T McInnes [2]Dayanjan S. Wijesinghe, Ph.D.

[1] Department of Clinical Practice, College of Pharmacy, Northern Border University, Rafha, Saudi Arabia

[2] Department of Pharmacotherapy and Outcomes Science, School of Pharmacy, Virginia Commonwealth University, Richmond, VA 23298, USA

[3] Department of Pharmacy Practice, Faculty of Pharmacy, University of Tabuk, Tabuk, Saudi Arabia

[4] Department of Clinical Practice, College of Pharmacy, Imam Abdulrahman Bin Faisal University, Dammam, Saudi Arabia

[5] Department of Pharmacy Practice, Unaizah College of Pharmacy, Qassim University, Unaizah, Saudi Arabia.

[6] Department of Clinical Pharmacy, School of Pharmacy, University of Hail 55476, Hail, Kingdom of Saudi Arabia

[7] Department of Computer Science, College of Engineering, Virginia Commonwealth University, Richmond, VA 23284, USA

**Abstract:**

The exponential growth of scientific literature has resulted in information overload, presenting significant challenges for researchers attempting to navigate and effectively synthesize relevant information from a vast array of publications. In this paper, we explore the potential of merging traditional reference management software with advanced computational techniques, specifically

Large Language Models (LLMs) and Retrieval-Augmented Generation (RAG), to address these challenges. We introduce PyZoBot, an AI-driven platform developed using Python that incorporates Zotero's reference management capabilities alongside OpenAI's sophisticated LLMs. PyZoBot is designed to streamline the extraction and synthesis of knowledge from extensive human curated scientific literature databases. Our work showcases PyZoBot's proficiency in handling complex natural language queries, integrating and synthesizing data from multiple sources, and meticulously presenting references to uphold research integrity and facilitate further exploration. By harnessing the combined power of LLMs, RAG, and the expertise of human researchers through a curated library of pertinent scientific literature, PyZoBot offers an effective solution to manage the deluge of information and keep pace with rapid scientific advancements. The development and implementation of such AI-enhanced tools promise to significantly improve the efficiency and effectiveness of research processes across various disciplines. The code for this work is publicly available at https://github.com/dayanjan/PyZoBot-RAG and permanently archived at DOI 10.5281/zenodo.16988871

**Key Points:**

- PyZoBot represents an innovative platform that integrates Zotero reference management with advanced RAG technology to address information overload in scientific research.
- The system offers four distinct implementations: standard RAG and knowledge graph-enhanced RAG, each available with either OpenAI models (GPT-3.5-Turbo, GPT-4o) or open-source models (LLaMA 3.1, Mistral).
- PyZoBot employs multiple document chunking strategies (Recursive, Layout-aware, and Semantic) to optimize text processing for different needs.

- Users can configure settings through an intuitive Streamlit interface, including Zotero API integration, model selection, embedding techniques, and retrieval parameters.
- The platform displays the specific document chunks used to generate responses, along with automatically generated citations and references, ensuring transparent and traceable information synthesis.
- Graph-RAG implementations offer visual knowledge graph exploration, allowing researchers to understand relationships between entities in their literature collections.
- PyZoBot supports both cloud-based deployments with OpenAI and fully local, privacy-conscious implementations using open-source models through Ollama.
- This implementation addresses limitations of traditional LLMs by grounding responses in curated literature and leveraging human expertise through organized Zotero libraries.
- PyZoBot serves as a comprehensive solution that demonstrates how AI-enhanced tools can improve research efficiency while maintaining scholarly integrity.

## 1. Introduction:

Information overload has become a prevalent problem in today's digital age, as individuals and organizations struggle to keep pace with the exponential growth of data and the constant influx of information from various sources[1]. While the rapid advancement of knowledge and communication technologies has made it easier to access and share information, it has also contributed to the overwhelming volume that people must process and manage[2]. This issue is particularly pronounced in academic research, where the amount of published literature has been growing rapidly[3].

In the biomedical field especially, over one million papers are published annually, making it increasingly difficult for researchers to stay updated with the latest developments[4,5]. The vast influx of scientific information not only overwhelms individual researchers but also hinders the ability to effectively identify meaningful trends and significant innovations within the broader scientific landscape[4,5]. Databases like PubMed are expanding at unprecedented rates, with thousands of new studies added daily, leading to significant challenges in navigating and filtering relevant information[5,6]. Consequently, researchers increasingly rely on information aggregators, automated alerts, and collaborative strategies to manage the influx of scientific publications, yet these coping mechanisms often demand considerable time and effort[6]. This explosion in publications contributes to "information overload," where the sheer volume of available data hinders efficient decision-making and slows the advancement of research[6].

Moreover, researchers conducting literature reviews face the challenge of navigating vast repositories to identify, evaluate, and synthesize relevant information[7], often spending significant time sifting through irrelevant, minimally relevant, or low-quality sources[8]. They may struggle to keep up with the latest developments, overlook important studies, or fail to identify critical gaps and opportunities[9]. The multidisciplinary nature of many research topics further compounds the problem, as researchers must integrate knowledge from various fields and disciplines, each with its own terminologies, methodologies, and publication practices[10,11], making it more challenging to identify and synthesize relevant information[12].

Large Language Models (LLMs) have emerged as a significant tool to tackle information overload, particularly in literature reviews and research synthesis across domains like biomedicine and social sciences[6,13,14]. LLMs such as GPT-4 and Bard offer researchers the ability to process vast amounts of text data quickly, identifying relevant studies, extracting key findings, and generating concise summaries of complex scientific literature[13,15]. By automating these traditionally labor-intensive tasks, LLMs significantly reduce the time required for conducting comprehensive reviews, allowing researchers to focus on deeper analysis and synthesis of the information[10]. For instance, LLMs can support the generation of hypotheses by identifying patterns and connections in the literature that may not be immediately apparent to human researchers[13]. Additionally, the ability of LLMs to cross-analyze texts from various domains facilitates interdisciplinary discoveries and highlights potential gaps in the literature[13,15]. However, it is essential to address challenges such as misinformation, bias, and ethical considerations, particularly as LLMs may inadvertently produce misleading outputs if not rigorously validated[13,15]. Despite these concerns, LLMs hold immense promise for transforming the way literature reviews are conducted, making research processes more efficient and innovative[10,15].However, despite their potential benefits, LLMs also have several limitations that should be considered when applying them to literature reviews. One major concern is the potential for biases and inaccuracies in the output generated. LLMs are trained on large datasets, which may contain biases, errors, or outdated information, leading to the risk of perpetuating these flaws in the summaries, syntheses, or recommendations produced by the model[15,16]. This can result in skewed conclusions or misleading directions in research, particularly when the LLM's training data lacks diversity or is sourced from unreliable material[17,18]. Another significant limitation is the tendency of LLMs to produce hallucinations or factually incorrect content when they generate text based on probabilistic models, rather than grounded knowledge, which can severely compromise the integrity of a literature review[18,19].

Additionally, LLMs have limited domain-specific knowledge and reasoning capabilities, which hampers their ability to grasp the nuances and complexities of specialized scientific fields. For instance, when faced with highly technical or context-specific literature, LLMs may oversimplify, omit critical details, or misinterpret information, leading to shallow or incomplete reviews[17,20,21]. Moreover, their lack of transparency and interpretability compounds these issues, as it becomes challenging for researchers to trace the reasoning behind the LLM-generated outputs, raising concerns about trust and accountability in academic research[13,18,22]. These limitations underscore the need for human oversight and rigorous validation when using LLMs in research contexts to mitigate potential errors and ensure the reliability of the results[16,18].

Retrieval-augmented generation (RAG) has emerged as a promising approach to address the limitations of LLMs in managing information overload and improving the quality of literature reviews. By combining the strengths of LLMs in generating coherent and fluent text with the ability to retrieve relevant, up-to-date information from external knowledge sources, RAG offers a significant improvement over traditional LLMs, particularly in avoiding hallucinations and ensuring factual accuracy[23–25]. One of the key advantages of RAG is its ability to retrieve and incorporate the most pertinent and current information from curated corpora, ensuring that the generated content is based on reliable and updated sources. In the context of literature reviews, this ability to dynamically access the latest research findings, methodologies, and expert insights from various databases or domain-specific repositories allows RAG models to produce well-informed and accurate summaries, which significantly reduces the cognitive burden on researchers[23,25].

Furthermore, RAG models can be tailored to retrieve information from specialized sources such as citation networks, expert-curated ontologies, or scientific databases, thereby enabling the models to capture the intricate details and contextual meanings of complex scientific literature[23,25]. This customization ensures that RAG can outperform traditional LLMs in understanding domain-specific nuances, making it a valuable tool for producing high-quality literature reviews. However, the full potential of RAG can only be realized if researchers carefully design and implement these systems, paying particular attention to selecting appropriate retrieval corpora and building domain-specific knowledge bases[26]. Additionally, human feedback and oversight should be integrated into the generation process to ensure that the output remains accurate and aligned with expert knowledge[24,25,27,28]. By combining the strengths of RAG-enhanced LLMs with human expertise, researchers can effectively manage the information overload associated with large-scale research endeavors.

While RAG has shown promise in addressing LLM limitations, significant challenges remain in implementing effective RAG systems for literature reviews. Researchers still face difficulties in accessing and maintaining up-to-date knowledge bases, ensuring reliable citation tracking, managing personal research collections effectively, and integrating RAG with existing research workflows and tools[25].These challenges highlight the need for specialized tools that can bridge the gap between RAG capabilities and researchers' practical needs.

One powerful external knowledge source that can be leveraged by RAG models is Zotero, a popular reference management software that allows researchers to collect, organize, and share bibliographic data, including articles, books, and other sources[29–31] . By integrating Zotero with RAG models, researchers can access a vast repository of curated and annotated scientific literature, enabling the models to retrieve the most relevant and up-to-date information for a given research topic. This approach can potentially mitigate biases, inaccuracies, and lack of domain-specific knowledge in traditional LLMs, leading to more accurate, comprehensive, and relevant summaries, syntheses, and recommendations. Moreover, using Zotero provides greater transparency by explicitly retrieving and citing sources, offering deeper insight into the reasoning and evidence behind the generated content, facilitating critical evaluation and validation of the output's quality, trustworthiness, and relevance to specific research objectives.

This article provides an in-depth examination of PyZoBot’s architecture, core components, and operational mechanisms. It details the implementation of RAG systems using both proprietary and open-source technologies. Specifically, it covers the development of RAG pipelines with vector store search powered by OpenAI models, as well as the integration of Knowledge Graphs to enhance retrieval performance. In parallel, it outlines analogous implementations utilizing open-source large language models. Collectively, this article delivers a thorough technical overview to facilitate a clear understanding of PyZoBot’s internal framework and capabilities.

## 2. Methodology for Implementing a RAG Pipeline: PyZoBot

### A- Software and Tools Used:

**1- Python:** Python (version 3.11.2) is a versatile and widely-used programming language known for its simplicity, efficiency, and object-oriented approach[32,33]. It is an interpreted language with dynamic typing and high-level data structures, making it ideal for various applications across different platforms. Python's popularity in fields like data science, machine learning, analytics, and geoprocessing is attributed to its robust standard libraries and ease of use. Moreover, Python's flexibility, visualization capabilities, and extensive library support make it a preferred choice for this project.

**2- Zotero:** Zotero, an open-source reference management software[30], is highly valued by a wide range of academic and professional users for its ability to simplify the collection, organization, and citation of research materials[29]. Developed at George Mason University, it plays a significant role in enhancing scholarly research and writing by streamlining the management of references, citations, and bibliographies[30]. One of its key features is the easy collection of references from various sources such as websites and academic journals, with automatic extraction of citation information from web pages and PDFs[23,34]. The user-friendly interface allows for the organization of references through folders, tags, and notes, ensuring quick retrieval. Notably, Zotero excels in generating citations and bibliographies in different styles like APA and MLA, thereby saving time on formatting[23,31]. Its integration with popular word processors like Microsoft Word and Google Docs enables users to directly insert citations and generate bibliographies in documents, ensuring accuracy and consistency[31]. Furthermore, its PDF management capability allows users to attach, organize, and annotate PDFs within the reference library[34]. Zotero fosters collaborative research through shared library features, which are essential for research teams[34]. Additionally, it offers cloud synchronization for easy access across devices and data backup, which enhances data security[35]. Browser extensions for Chrome and Firefox simplify the process of capturing online references[23,36]. As an open-source software, Zotero is continuously improved through community contributions, and its availability on multiple platforms expands its user base. Its applications are diverse, benefiting academic research, education, library services, as well as professionals in fields such as legal, medical, and media, by facilitating the management and citation of a wide range of references.

**3- Ollama:** Ollama is an open-source framework designed to simplify the deployment and utilization of LLMs in local environments[37]. It enables users to run, customize, and interact with a variety of open-source LLMs such as Llama and Mistral directly on their personal computers without requiring specialized hardware or cloud infrastructure[38,39]. This tool provides a streamlined command-line interface and API that allows for easy model management, including downloading, running, and customizing models through a process called "Modelfiles" [67].

For research purposes, Ollama offers significant benefits, including data privacy (as all processing occurs locally without data leaving the user's device), cost efficiency (eliminating cloud service fees), and full control over model customization[40]. In the context of PyZoBot, Ollama serves as the runtime environment for deploying open-source large language models. This enables the RAG pipeline to function entirely locally when needed, providing an alternative to cloud-based LLM APIs like OpenAI. The integration of Ollama allows PyZoBot to maintain flexibility in its deployment options, giving users the choice between proprietary cloud models (like GPT-4) or open-source alternatives (like LLaMA 3 or Mistral) depending on their specific requirements for privacy, cost, or performance. Furthermore, Ollama's ability to operate offline[41] ensures that PyZoBot can process and analyze sensitive research materials without internet connectivity, which is particularly valuable in environments with restricted network access or when working with confidential data.

For scholarly applications, Ollama represents a significant advancement in democratizing access to LLM capabilities, allowing researchers to harness AI-powered knowledge extraction from their curated Zotero libraries without relying exclusively on commercial API services. This aligns with PyZoBot's goal of creating accessible tools that enhance research productivity while respecting data privacy and providing sustainable, cost-effective solutions for academic work.

### B- Build RAG system with vector store search:

This section describes the technical implementation of PyZoBot's Retrieval-Augmented Generation system, which integrates Zotero library data, document chunking, embedding generation, vector store indexing, similarity search, and LLM generation. The architecture follows the pipeline illustrated in the workflow diagram (Figure 1), combining retrieval and generation components to enhance the accuracy and transparency of responses generated from scientific literature.

#### Step 1: Data Source - Zotero Library Access

PyZoBot begins by retrieving documents from a Zotero reference library. Users provide their Zotero API key, library ID, and optionally a collection ID through the Streamlit interface. The function `process_zotero_library()` (in both OpenAI and OpenSource versions) connects to Zotero using the PyZotero library from python, fetches item metadata, and downloads associated PDFs.

```
def process_zotero_library(library_id, zotero_api_key, library_type,
collection_id=None):
    zot = zotero.Zotero(
        library_id=library_id, library_type=library_type, api_key=zotero_api_key
    )
    items = zot.everything(zot.top())
```

```
# Process Zotero items and Extract PDFs...

df = pd.json_normalize(items)
```

**Step 2: Document Chunking**

Once PDFs are retrieved, they are processed into smaller text chunks. PyZoBot supports three chunking methods (table 1):

- **Recursive Chunking**: Recursive chunking breaks down documents into fixed-size overlapping text segments[42,43], splitting the text using the `RecursiveCharacterTextSplitter` component[44]. This process begins by attempting to split the text using a prioritized list of separators, such as ["\n\n ", "\n ", ""][44]. If the resulting segments still exceed the defined chunk size, the splitter recursively applies the next separator, progressively refining the chunks until they fit within the desired length[44]. This recursive process offers flexibility, allowing PyZoBot to preserve meaningful structural elements from the document, especially when dealing with complex academic papers that contain mixed formatting styles[45]. While recursive chunking may require more processing time and computational resources compared to simpler fixed-length segmentation, it enhances the relevance of retrieved content by maintaining stronger contextual continuity between chunks[46]. This makes recursive chunking particularly useful for processing lengthy scientific papers, generating concise summaries, and supporting retrieval in knowledge bases containing extensive technical documentation or large sets of FAQs[46].

- **Layout-aware chunking**: The layout-aware chunking strategy, implemented using the unstructured library's `partition_pdf` function, preserves the document's original structure and layout during segmentation[47,48]. This method is particularly valuable when dealing with documents where formatting and structural integrity are essential.
- **Semantic Chunking**: Semantic chunking is an advanced technique that divides a document into meaningful units based on the semantic relationships present in the text. Unlike traditional methods that rely on fixed sizes or simple delimiters, semantic chunking identifies boundaries based on the underlying meaning and context, ensuring each chunk is a coherent and complete semantic unit[46,49]. This approach typically utilizes pre-trained language models to generate text embeddings that capture the semantic essence of the content. By clustering or grouping sentences with similar embeddings, the system ensures that semantically related information stays together, enhancing retrieval accuracy[49]. This method is especially beneficial for processing complex documents with intricate structures, such as legal texts or financial reports, where maintaining the contextual relationship between different sections is essential[50]. Furthermore, by transforming documents into semantically coherent chunks, the retrieval process becomes more precise and contributes directly to the improved performance of RAG systems[51]. However, semantic chunking can be computationally intensive and requires careful tuning of parameters, such as embedding models and similarity thresholds, to match the specific nature of the documents being processed[46].

In PyZoBot, semantic chunking uses the "gradient" method to identify chunk boundaries. This method analyzes the gradient of embedding distances between consecutive text segments and applies anomaly detection techniques to identify points where the semantic flow shifts significantly. This is particularly useful for domain-specific documents, such as biomedical or scientific literature, where closely related content needs to be kept intact within a chunk[52].

**Table 1: Comparison for Recursive Chunking, Layout-aware Chunking, and Semantic Chunking.**

| Feature/Aspect | Recursive Chunking | Layout-aware Chunking | Semantic Chunking |
|---|---|---|---|
| **Definition** | Divides text hierarchically using separators like paragraphs or sections[49]. | Respects document layout and formatting, preserving sections, headers, and tables[53]. | Segments text based on semantic relationships, maintaining contextual integrity[45]. |
| **Focus** | Hierarchical structure and optimal chunk size[46]. | Document's inherent organization and spatial consistency[49]. | Semantic coherence and meaning[54]. |
| **Advantages** | Adapts to document structure, ensuring chunks are not cut mid-sentence[49]. | Maintains layout consistency, preserving document meaning[53]. | Enhances retrieval and analysis by preserving semantic relationships[46]. |
| **Challenges** | May require multiple iterations to achieve desired chunk size[46]. | May not capture semantic relationships effectively[49]. | Computationally intensive and may not always justify the overhead[54]. |

### Step 3: Embedding Creation

Each chunk is converted into a dense vector embedding that captures its semantic meaning. The embedding model depends on the implementation version:

- **OpenAI Version (PyZoBot_OpenAI):** PyZoBot utilizes OpenAI's "text-embedding-3-large" model to transform textual content into numerical vectors that capture semantic meaning. This embedding model plays a crucial role in the RAG process by acting as a dense retriever that enhances the performance of LLMs. Known for its strong performance across various retrieval tasks, it excels at retrieving document segments that preserve essential context, especially in lengthy documents[55,56]. In the PyZoBot pipeline, the *text-embedding-3-large* model is directly integrated into the retrieval process to enhance accuracy and reduce the risk of hallucinated content, which is particularly valuable in specialized fields such as healthcare.
- **Open-Source Version (PyZoBot_OpenSource):** In developing the open-source version of PyZoBot, referred to as *PyZoBot_OpenSource*, careful consideration was given to selecting suitable embedding models that could support biomedical question answering tasks. The selection process was guided by two main criteria: first, the model needed to be available through Ollama, a platform that simplifies running and managing large language models; second, the model needed to handle input sequences of at least 8191 tokens to support processing longer biomedical texts. Based on these criteria, the *Nomic* embedding model was selected, as it supports sequences up to 8192 tokens[57,58]. Additionally, *BGE-M3*, part of the BGE (BAAI General Embedding) series developed by the Beijing Academy of Artificial Intelligence (BAAI), was also incorporated[59,60]. BGE-M3 is designed to handle a wide range of text lengths, from short sentences to long documents, and it also supports up to 8192 tokens[60]. Both models were fully integrated into *PyZoBot_OpenSource*, and the system was designed to allow easy modification, so future models can be added if needed. This flexible and open approach enables *PyZoBot_OpenSource* to adapt to future advancements by incorporating new and improved embedding models as they become available, ensuring ongoing innovation in biomedical natural language processing.

**Step 4: Vector Store Indexing**

In PyZoBot, vector store indexing is implemented using ChromaDB, an open-source database specifically designed for storing and utilizing vector embeddings along with their relevant metadata. ChromaDB is particularly effective in applications requiring efficient retrieval and comparison of high-dimensional data, such as those involving LLMs and machine learning tasks[61,62]. After documents are chunked and their embeddings are generated, PyZoBot stores these embeddings in ChromaDB, creating a persistent vector index that supports fast retrieval during the query process. The embeddings are stored alongside metadata such as the source document name and the position of each chunk within the document, allowing for transparent citation in generated responses. ChromaDB's reliance on cosine similarity ensures efficient and accurate querying of semantically relevant document chunks, which is essential for RAG systems[62]. Furthermore, its integration into pipelines allows seamless handling of large-scale document collections, including biomedical datasets, supporting advanced processing tasks like clinical text analysis and ontology term prediction[62,63]. This makes ChromaDB an ideal choice for PyZoBot's.

**Step 5-8: User Query Processing**

When the user submits a query through the Streamlit interface, it is converted into a vector embedding using the same embedding model (OpenAI or Open source). This ensures the query exists in the same vector space as the document chunks.

**Step 9-10: Vector Similarity Search and Information Retrieval**

Once the user query is embedded, the system performs a similarity search using ChromaDB to retrieve the *k* most relevant text chunks (***k*** *is a parameter that represents the number of the most relevant items text chunks)*. To enhance the retrieval process, PyZoBot applies Maximum Marginal Relevance (MMR), a technique designed to balance relevance and diversity in the retrieved results[64–66]. In typical `LangChain` implementations of MMR, a parameter known as *lambda* (sometimes referred to as `lambda_mult` or simply *lambda*) controls this balance. Specifically, *lambda* determines the trade-off between relevance, which reflects how closely a document matches the query, and diversity, which ensures that the selected documents are not too similar to each other. In many existing implementations, the commonly used default value for *lambda* is 0.5[67,68] , which has also been adopted as the default in PyZoBot.

```
compressor = LLMChainExtractor.from_llm(llm)

compression_retriever = ContextualCompressionRetriever(

    base_compressor=compressor,

    base_retriever=db.as_retriever(

    search_type="mmr", search_kwargs={"k": st.session_state["k"], "lambda_mult":
0.5}

        ),

    )
```

**Step 11: Augmented Prompt**

In RAG, the prompt serves as the essential bridge between retrieval and generation processes. A prompt in this context is not simply a user query; rather, it is a structured input that combines the user's question with relevant contextual information retrieved from an external knowledge source, such as a vector database[24,69]. The purpose of the prompt is to provide the LLM with all the necessary background information to generate a complete, contextually accurate, and traceable response. This is especially valuable in domains like scientific research and biomedicine, where factual precision is crucial[70–73]. The integration of relevant retrieved content into the prompt ensures that the LLM's response does not rely solely on its pre-trained parametric knowledge, which can become outdated or incomplete[71,73,74].

In PyZoBot, the prompt is designed to follow a structured evidence-based format that maximizes accuracy, transparency, and traceability. The retrieved chunks (top-K results from ChromaDB) are combined into a prompt alongside the user's query, ensuring that every generated response is grounded in the retrieved content.

```
system_template = """
   Answer the user query using ONLY the relevant content provided in this prompt.
   Do not use any external knowledge or information does not present in the given
context.
   If the provided content does not contain information to answer the query,
respond with "I don't have enough information to answer this question based on
the given context."
   Take your time and provide as much information as you can in the answer.\n

   For each statement in your answer, provide in-text citations after the
sentence, e.g., [1].
   Start with number [1] every time you generate an answer and make the number
matching the source document.
   If a statement has multiple citations, provide them all, e.g., [1], [2], [3].

   By the end of the answer, provide a References section as Markdown (###
References) including the number and the file name, e.g.:
   [1] Author et al. - YEAR - file name.pdf

   Write each reference on a new line.

    {summaries}
```

```
"""
```

**Step 12-13: Response Generation and Delivery**

Once the augmented prompt is fully constructed, it is passed to the LLM for final response generation. The LLM selection depends on the implementation version:

- **OpenAI Version (PyZoBot_OpenAI):** In the OpenAI-based implementation of PyZoBot, the system relies on OpenAI's GPT models, specifically GPT-4o, or GPT-3.5-turbo, to generate responses based on the augmented prompt. After the retrieval and prompt construction stages are completed, the augmented prompt is passed directly to the chosen OpenAI model using the `ChatOpenAI` interface from LangChain. This integration allows PyZoBot to leverage OpenAI's state-of-the-art language understanding and generation capabilities. The system dynamically selects the appropriate OpenAI model based on user preference, offering flexibility between cost (GPT-3.5-turbo) and performance (GPT-4o). This approach allows PyZoBot to operate effectively in cloud-connected environments where access to OpenAI's API is available.
- **Open-Source Version (PyZoBot_OpenSource):** In the open-source implementation of PyZoBot, the system uses locally hosted language models via the Ollama platform. Specifically, PyZoBot supports Mistral-7B and LLaMA 3.1 (Llama-3.1-8B), two widely recognized open-source models with strong performance across various natural language processing tasks[69,75]. After the augmented prompt is prepared, it is passed to the selected local model through the `ChatOllama` interface. This implementation allows PyZoBot to operate entirely offline, making it suitable for privacy-sensitive environments or research setups where internet access is restricted.

An important parameter in answer generation is the model temperature. In LLM, temperature controls the degree of randomness in the model's output. A higher temperature (e.g., 0.7 or 1.0) encourages the model to explore less probable word choices, leading to more creative and diverse responses. However, this increased creativity comes at the cost of reduced consistency and greater variability, meaning the model's output can become unpredictable and may drift away from factual accuracy[76,77]. In contrast, setting the temperature to 0 makes the model fully controlled. In this mode, the model always selects the highest-probability token at each generation step, ensuring the response is consistent and repeatable if the same input is provided. This setting is particularly valuable in factual, evidence-based applications such as RAG systems, where accuracy and reliability are more important than creativity[77,78]. In PyZoBot, the temperature is explicitly set to 0 for all generated responses. This decision ensures that every answer provided by the system is fully grounded in the retrieved content from the Zotero library and is not influenced by unsupported or speculative tendencies.

Once the response is generated, it is sent directly to the Streamlit interface, where the user sees the final answer. To enhance transparency, the response includes in-text citations within the generated text, linking each statement back to the corresponding retrieved document chunk. At the end of each response, PyZoBot automatically generates a References section, formatted in Markdown, listing the cited documents by file name. This design ensures that PyZoBot's outputs are fully traceable, enabling users to verify each statement and reference the original document if further review is required. This feature is especially critical in academic and biomedical contexts, where transparency and evidence tracking are essential.

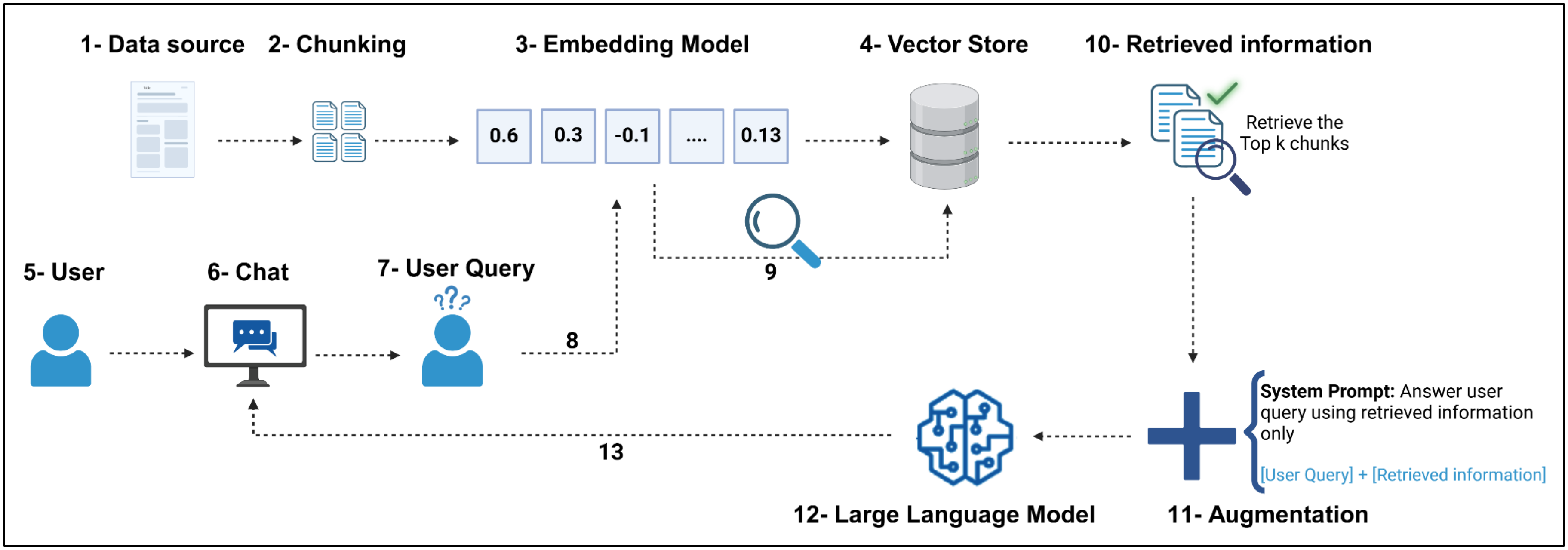

**Figure 1: PyZoBot- RAG pipeline**

This diagram illustrates the workflow of a Retrieval-Augmented Generation (RAG) system for answering user queries using a LLM. The process begins with (1) a data source, such as documents (PDFs retrieved from Zotero library), which is (2) chunked into smaller segments. These chunks are then (3) processed by an embedding model, converting them into vector representations. These vectors are stored in (4) a vector store for efficient retrieval. When (5) a user interacts with (6) a chat interface and submits (7) a user query, the query is (8) embedded using the same embedding model. The embedded query is used to (9) search the vector store, (10) retrieving the top-k most relevant chunks. The retrieved information is then (11) augmented with the user query, combining both into a prompt for the (12) large language model. The system prompt ensures that the LLM answers the user query using only the retrieved information. Finally, (13) the generated response is returned to the user through the chat interface.

### C- Build Graph-RAG system:

RAG has emerged as a powerful approach to enhance LLMs by integrating external knowledge sources. Traditional RAG implementations rely on vector-based retrieval, which, while effective, can struggle with complex relationships within unstructured data. Graph Retrieval-Augmented Generation (Graph-RAG) extends this capability by using knowledge graphs to structure and retrieve information more effectively. This approach is particularly advantageous in biomedical research, where relationships between entities (e.g., genes, diseases, drugs) play a crucial role in knowledge synthesis[79–81].

The Graph-RAG system developed in this dissertation integrates a knowledge graph-based retrieval framework with LLMs, enabling enhanced comprehension and contextual reasoning. The pipeline consists of the following stages (Figure 2):

**Step 1: Data Source - Zotero Library Access**

This step has been previously discussed in Step 1 of RAG in Section B.

**Step 2: Document Chunking**

This step has been previously discussed in Step 2 of RAG in Section B.

**Step 3: Entity and Relationship Extraction**

In our Graph-RAG system, we employ LlamaIndex's `PropertyGraphIndex` to construct a structured knowledge graph from text chunks, enabling efficient querying and information retrieval. A property graph is a specialized type of graph database where nodes (entities) and edges (relationships) possess associated attributes, allowing for a richer representation of data and facilitating complex queries[82–85]. By leveraging this index, our system effectively models relationships within the document corpus, thereby enhancing tasks such as relationship discovery, knowledge representation, and downstream analytical processes in fields such as natural language processing and knowledge graph construction. The implementation of a property graph index is particularly advantageous in scenarios where understanding the relationships between textual entities is critical, such as in information retrieval, question answering, and knowledge base population.

To construct the knowledge graph, our Graph-RAG system employs two distinct extractors:

- ImplicitPathExtractor: is a component within the LlamaIndex framework designed to construct knowledge graphs by establishing implicit relationships between text chunks. Unlike methods that rely on LLMs to extract explicit entity relationships, the `ImplicitPathExtractor` operates without LLMs or embedding models[82,85].
- SimpleLLMPathExtractor: the `SimpleLLMPathExtractor` takes a more explicit approach by utilizing a large language model to identify relational paths between entities in the text[82]. Depending on the implementation, different LLMs are employed: in the OpenAI-based variant, we leverage GPT-4o or GPT-3.5-turbo, while in the open-source version, models such as Mistral and LLaMA-3.1-8B from Ollama are utilized. To ensure computational efficiency while maintaining the depth of extracted relationships, we configure the extractor to retrieve up to five relational paths per text chunk. We restrict the number of extracted paths to five per text chunk to regulate the number of relationships identified, ensuring a balance between capturing essential connections and maintaining computational efficiency.

```
index = PropertyGraphIndex.from_documents(
    documents=llama_docs,
    embed_model=embed_model,
    kg_extractors=[
        ImplicitPathExtractor(),
        SimpleLLMPathExtractor(
```

```
                    llm=llm,

                    num_workers=2,

                    max_paths_per_chunk=5)
```

**Step 4: Embedding Creation**

In LlamaIndex's `PropertyGraphIndex`, an embedding model is employed to generate vector representations for graph nodes, enabling efficient search and retrieval[85]. In our Graph-RAG implementation, we utilized the same embedding models for both the OpenAI and open-source versions (Section B, Step 3).

**Step 5: Graph Construction and Indexing**

In the Graph Construction and Indexing step of the Graph-RAG pipeline, the extracted entities and relationships are organized into a structured knowledge graph, with nodes representing entities and edges capturing their relationships. This process is enriched by attaching embeddings—generated earlier by the embedding model—to each node and edge, enabling semantic similarity searches. The `PropertyGraphIndex.from_documents` function in `create_graph_index` performs this dual task: it constructs the graph using extractors like `ImplicitPathExtractor` and `SimpleLLMPathExtractor` to define nodes and edges, while simultaneously indexing the embeddings within the graph structure for efficient retrieval. This indexing ensures that the system can quickly search for relevant nodes and traverse relationships during query processing. The resulting `graph_index` becomes the backbone for subsequent retrieval and generation tasks, enabling the system to reason over structured knowledge and retrieve contextually relevant information.

**Step 6-9: User Query Processing**

In the **User → Chat → User Query** phase of the Graph-RAG pipeline, the process starts with the user interacting with the "Chat" interface, implemented using Streamlit library from python, where they input a natural language query via the `st.text_input` widget. Upon clicking the "Get Answer" button, the query is captured and sent to the `answer_question` function, where it is processed by the `query_engine` constructed from the `PropertyGraphIndex`. In this implementation, the `query_engine` operates in `retriever_mode="keyword"`, meaning it extracts keywords from the query (using a keyword extraction mechanism). These keywords are then used to search the graph index, enabling the system to retrieve relevant entities, relationships, and text chunks from the knowledge graph store, setting the stage for augmentation and response generation.

**Step 10: Retrieve Relevant Information**

In this step, the system leverages the keywords extracted from the user query to fetch relevent data from the knowledge graph store, marking a critical transition from search to actionable context. This keyword-based search identifies the most relevant entities, relationships, and text chunks based on the query terms. The retrieved information, captured in `response.source_nodes`, includes details such as source file names and text content, which are then compiled as chunks and graph data for the subsequent augmentation step, providing the foundation for generating a detailed response.

### Step 11: Augmentation

In the augmentation phase of the Graph-RAG pipeline, the system combines the user query with the relevant information retrieved from the knowledge graph store to prepare a rich context for response generation. After the `query_engine` retrieves text chunks, entities, and relationships based on keyword matches, this step integrates the user query with the retrieved data, such as text snippets and associated graph elements. The `custom_prompt` defined previously in (section B, step 9), ensuring the LLM receives the query alongside the context from `response.source_nodes`, which includes source file names and text content. The combined input, which includes the user query, retrieved text chunks, and graph data, is sent to the LLM, allowing it to create an answer using the keyword-matched results and the knowledge graph.

**Step 10-11: Response Generation and Delivery**

These steps have been previously discussed in Step 10-11 of RAG in Section B.

**Graph Visualization and the "Knowledge Graph" Tab**

The system generates an interactive visual representation of the knowledge graph, allowing users to explore relationships between entities directly. Once the knowledge graph is constructed and indexed in the `PropertyGraphIndex`, the `create_graph_visualization` function converts the graph into a NetworkX format and saves it as an HTML file (`kg_viz.html`) using the `save_networkx_graph` method. This visualization, displayed in the "Knowledge Graph" tab of the Streamlit interface, enables users to interact with the graph by dragging nodes, zooming, and examining connections, thereby providing an intuitive means of understanding the structured knowledge extracted from the documents. Additionally, the tab includes a download button that allows users to save the visualization, ensuring they can retain and revisit the interactive graph for future reference.

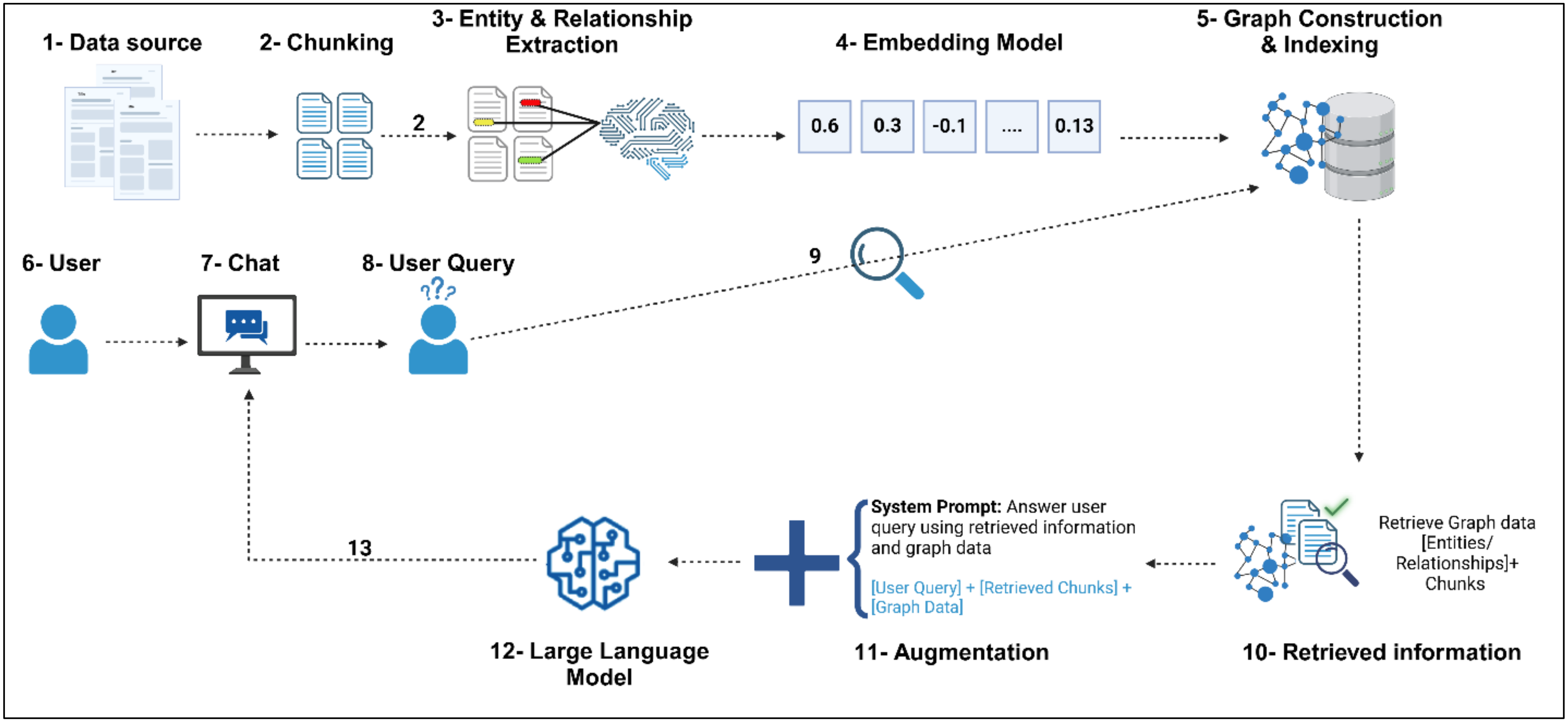

This figure illustrates the GRAPH-RAG (Graph-based Retrieval-Augmented Generation) pipeline, detailing the flow of data and processing steps from raw data to a generated response. The process begins with **1. Data Source**, where raw documents (PDFs) are collected, followed by **2. Chunking**, which breaks them into smaller text pieces. **3. Entity & Relationship Extraction** identifies key entities and their connections using an LLM or NLP techniques. **4. Embedding Model** converts text chunks, entities, and relationships into numerical vectors, while **5. Graph Construction & Indexing** builds and indexes a knowledge graph store with nodes (entities), edges (relationships), and embeddings. The user initiates the query process at **6. User**, interacting via **7. Chat**, leading to **8. User Query**, which is processed by the system. **9. Search** uses keyword matching to retrieve relevant data from the knowledge graph store, guided by the query. **10. Retrieved Information** includes graph data (entities, relationships, and chunks). **11. Augmentation** combines the user query, retrieved chunks, and graph data using a custom system prompt to guide the LLM. **12. Large Language Model** generates the response based on this augmented input, and **13. Chat** delivers the final answer back to the user.

## Results:

In this chapter, we present the results of implementing four distinct RAG approaches within the PyZoBot platform to facilitate efficient information synthesis from curated Zotero reference libraries. To enhance usability and accessibility, PyZoBot features a Graphical User Interface (GUI) built using Streamlit, providing an interactive environment for researchers. The system offers four implementations: two standard RAG systems and two knowledge graph-enhanced RAG (GraphRAG) systems, with both OpenAI-based and open-source variants.

### 1. User Interface and Interaction Design

The PyZoBot user interface (UI) is designed to be intuitive and researcher-friendly, allowing users to interact seamlessly. The platform follows a structured, multi-step workflow that systematically guides users through key processes, including Zotero integration, document processing, query execution, and information synthesis. This structured approach enhances usability while ensuring a smooth and efficient experience in literature exploration and synthesis.

#### 1.1 Home Screen and System Selection (figure 3)

Upon launching PyZoBot, users are presented with a centralized home screen where they can select one of the four system implementations:

- PyZoBot OpenAI (Standard RAG with OpenAI's LLMs)
- PyZoBot OpenSource (Standard RAG with Open-Source LLMs)
- PyZoBot GraphRAG OpenAI (Knowledge Graph-Enhanced RAG with OpenAI's LLMs)
- PyZoBot GraphRAG OpenSource (Knowledge Graph-Enhanced RAG with Open-Source LLMs)

Each implementation has a dedicated UI section, enabling users to configure its parameters without switching between multiple tools.

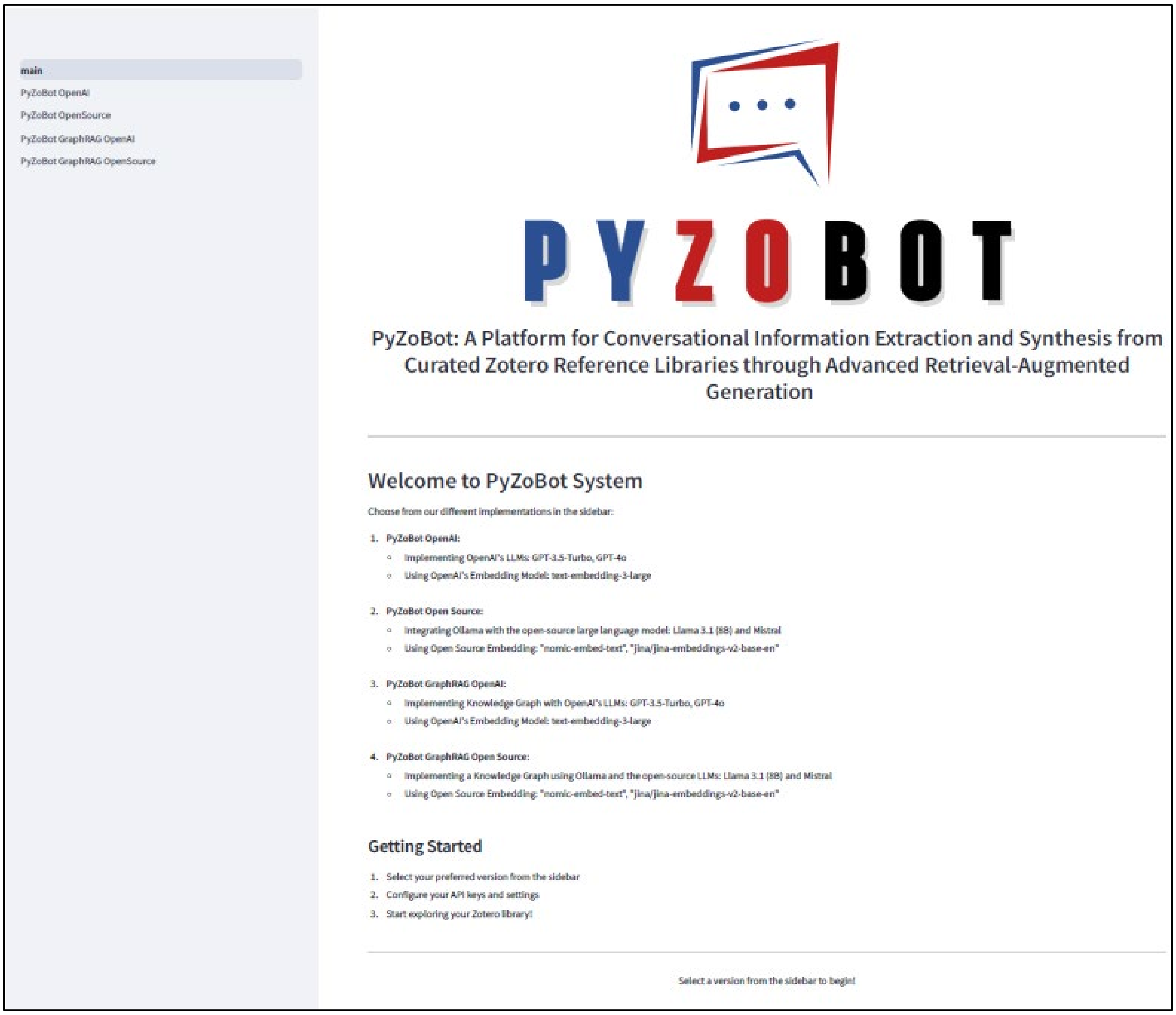

**Figure 3: Main Home Screen.**

### 2. Sidebar: Configuration and System Setup

The **left sidebar** serves as the control panel, allowing users to configure PyZoBot according to their needs. It is divided into multiple sections:

#### 2.1 Zotero Integration and Library Configuration

Users begin by inputting their Zotero API key to establish a secure connection with their Zotero library. They then select the library type, choosing between a group or user library, and specify the corresponding library ID. To refine document retrieval, users can optionally provide a collection ID, allowing them to target specific reference collections within Zotero. Once these details are entered, the "Fetch PDFs from Zotero" button initiates the process of retrieving metadata and full-text PDFs, ensuring that relevant documents are available for subsequent processing and analysis. The Zotero configuration process is illustrated in Figure 4.

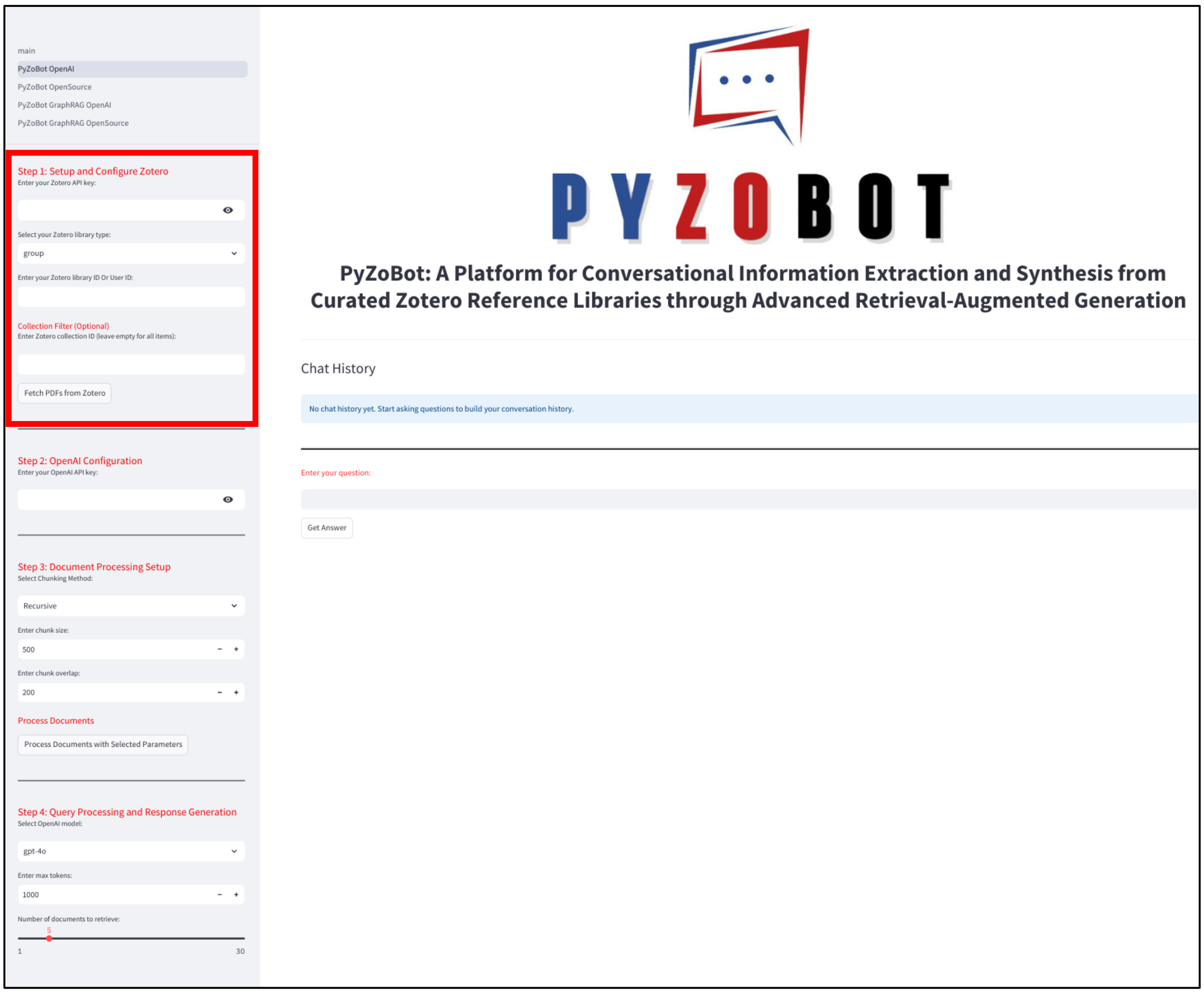

**Figure 4: Zotero Integration and Library Configuration.** This figure illustrates the initial setup process for integrating Zotero with PyZoBot. Users input their Zotero API key to connect securely, select a library type (group or user) and ID,

and optionally provide a collection ID to refine retrieval. The "Fetch PDFs from Zotero" button then retrieves metadata and full-text PDFs for further processing.

### 2.2 LLM Model Selection

Users have the option to select different LLMs based on their preferences and system requirements. For OpenAI-based implementations, they can choose between GPT-3.5-Turbo and GPT-4o (figure 5), both of which provide powerful natural language understanding and generation capabilities. Alternatively, for open-source implementations, users can opt for LLaMA 3.1 (8B) or Mistral (figure 6), which are deployed through Ollama. These open-source models allow for greater customization and local execution, ensuring flexibility and privacy-conscious deployments.

### 2.3 Embedding Model Configuration

In OpenAI-based versions, the system leverages text-embedding-3-large, a powerful embedding model designed for high-quality vector representations of textual data. This enables efficient retrieval of relevant literature by converting text into numerical embeddings that facilitate similarity search. For Open-Source implementations, users have the flexibility to choose between nomic-embed-text and bge-m3, both of which are optimized for open-source environments. These models allow PyZoBot to operate independently of proprietary APIs, ensuring greater control over data privacy and computational resources while maintaining robust retrieval capabilities.

### 2.4 Document Processing Settings

Users can select one of three text chunking methods to optimize document segmentation for retrieval and synthesis. Recursive Chunking, which is the default method, systematically divides text into smaller segments while preserving context. Layout-Aware Chunking maintains the structural integrity of the document, ensuring that formatting and visual elements are considered during text processing. Semantic Chunking utilizes AI-driven techniques to segment text based on meaningful content breaks, enhancing contextual relevance. Additionally, users can fine-tune retrieval quality by adjusting chunk size and overlap settings, allowing for greater control over how information is processed and indexed within the system.

### 2.5 Query and Response Parameters

Users have the ability to configure query processing by adjusting key response parameters. They can modify the maximum token limit for responses generated by the LLM, ensuring that answers are concise or detailed depending on their research needs. Additionally, users can define the number of top documents (k) to retrieve, which determines how many reference materials are considered in response generation. These settings allow for flexible customization, balancing comprehensiveness with efficiency in retrieving and synthesizing literature.

## 3. Main Interaction Panel: Query Execution and Literature Synthesis

### 3.1 Query Input and Response Display

The query input and response display section allows users to interact with PyZoBot through a simple text box where they can enter natural language queries related to their research. Upon receiving a query, the system retrieves the most relevant references from the indexed Zotero documents using its retrieval-augmented generation (RAG) pipeline. The retrieved references are then synthesized into a coherent response, ensuring that key insights are accurately represented. To maintain research integrity and transparency, PyZoBot cites the sources used in generating the response, enabling users to verify the information and explore the referenced literature further.

### 3.2 Knowledge Graph Visualization (GraphRAG Implementations Only)

For GraphRAG implementations, the UI dynamically renders a knowledge graph, visually representing relationships among retrieved entities. Users can interactively explore nodes and connections to gain a deeper understanding of how retrieved documents relate to one another. This feature enhances transparency and contextual comprehension by allowing users to navigate through structured representations of their reference materials, making it easier to identify connections, trends, and relevant sources within the dataset.

The user interfaces for both GraphRAG implementations—OpenAI-based and open-source—are visually distinct yet functionally aligned. Figure 7 presents the PyZoBot GraphRAG OpenAI interface, highlighting its dual-panel layout that includes a chat interaction tab and a dynamic knowledge graph visualization. Similarly, Figure 8 depicts the GraphRAG Open-Source interface, which mirrors the OpenAI version in structure while enabling interactions with locally hosted LLMs. Both interfaces provide configuration panels for Zotero integration, model selection, and chunking strategies, allowing users to customize their information retrieval workflows. Further, Figure 9 showcases a biomedical knowledge graph generated using the PyZoBot GraphRAG OpenAI implementation. This visualization allows users to trace relationships among biomedical entities.

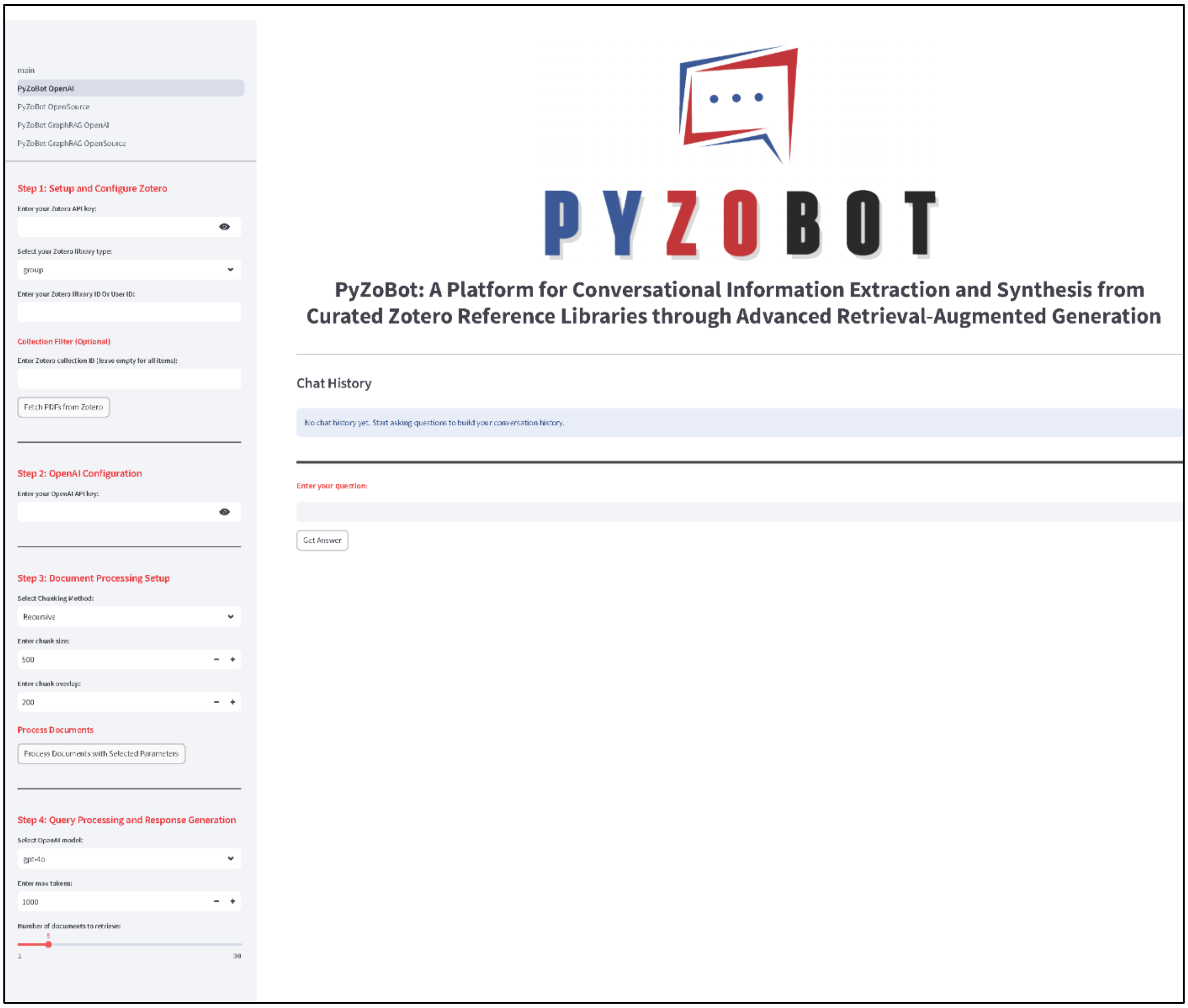

**Figure 5: OpenAI Integration and Configuration in PyZoBot.** This figure illustrates the setup process for integrating OpenAI with PyZoBot. Users input their OpenAI API key to establish a connection, select the desired model (gpt-4o or gpt-3.5-turbo), and configure parameters such as maximum tokens (Max= 4000) and the number of documents to retrieve (1 to 30). These settings enable PyZoBot to leverage OpenAI's language models for conversational information extraction and response generation from curated reference libraries.

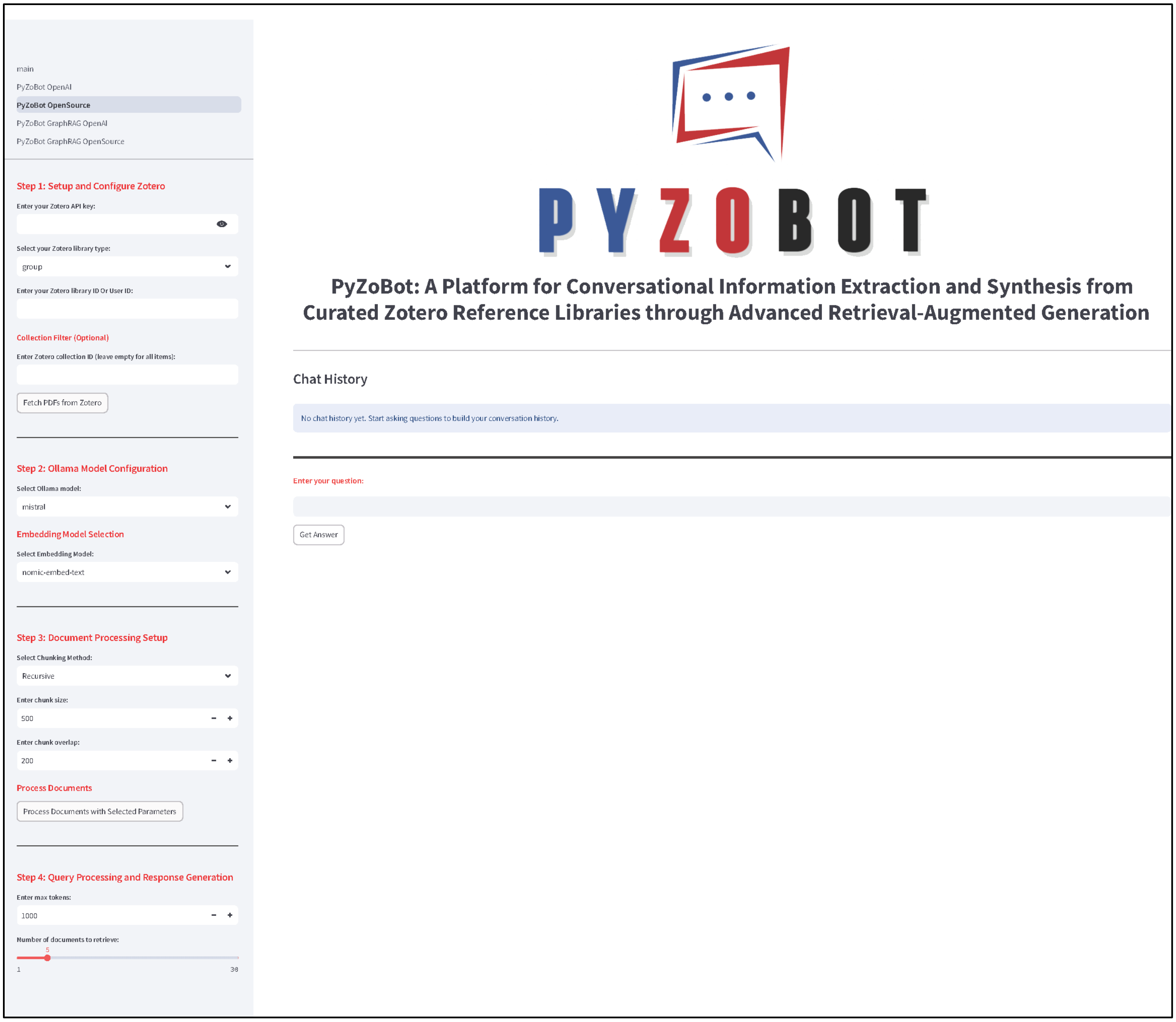

**Figure 6: OpenSource Configuration in PyZoBot.** This figure illustrates the document processing setup in PyZoBot, a platform for conversational information extraction and synthesis from curated Zotero reference libraries using advanced retrieval-augmented generation. Users connect to their Zotero library via an API key, and choose a model (mistral or LLaMA 3.1) with embedding models (e.g., nomic-embed-text or bge-3). The chunking setup allows users to select a method (recursive with customizable chunk sizes, layout-aware, semantic) for efficient document processing, while response generation parameters include maximum tokens (Max = 4000) and the number of documents to retrieve (1 to 30), enabling effective conversational interactions.

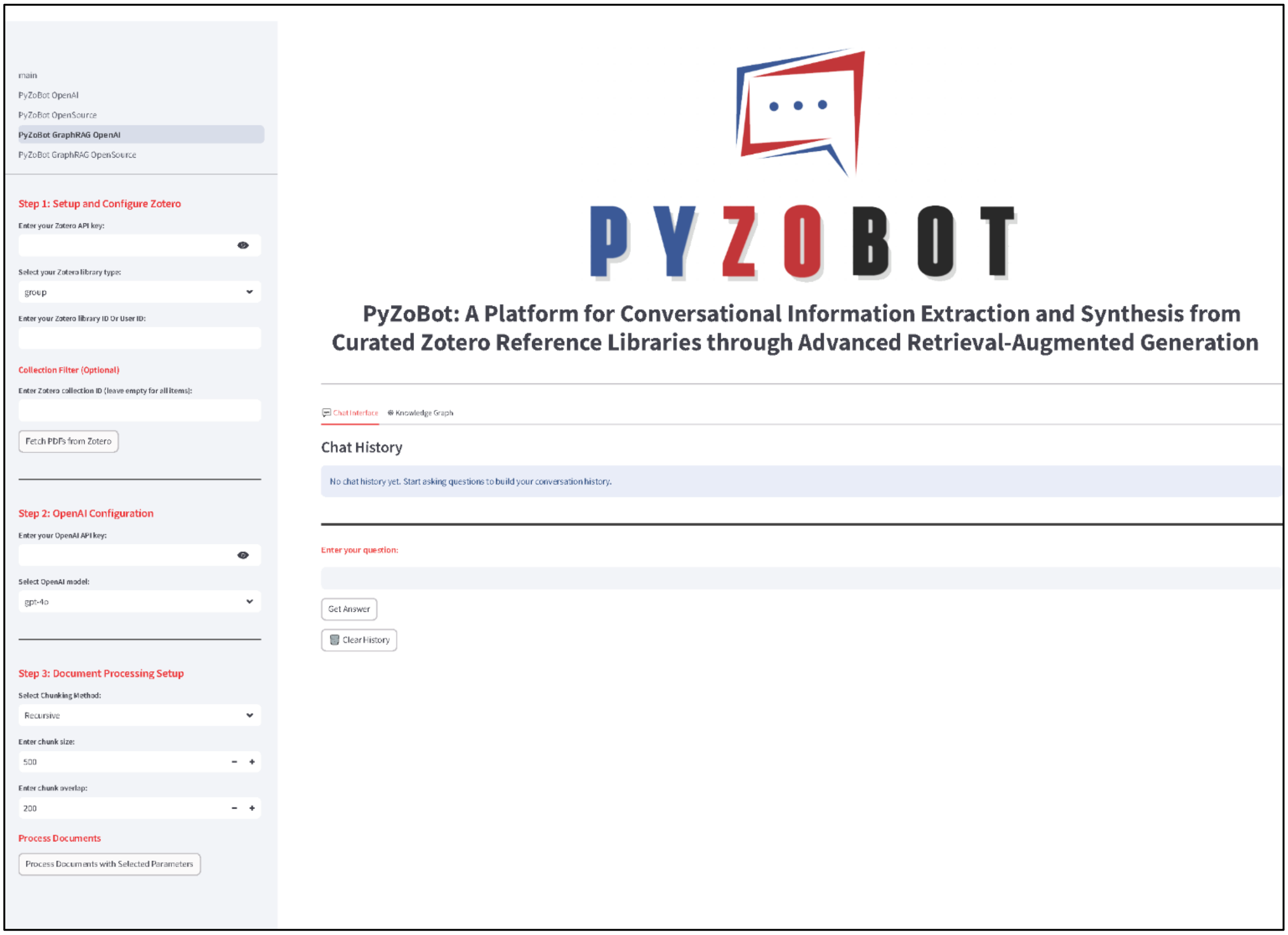

**Figure 7: PyZoBot GraphRAG OpenAI User Interface with Knowledge Graph.** This figure displays the PyZoBot GraphRAG OpenAI interface. The left sidebar provides configuration options, including Zotero API setup, OpenAI model selection (e.g., GPT-4o or GPT-3.5-Turbo), and document processing settings such as chunking methods. The query interaction panel is divided into two tabs: Chat Interface and Knowledge Graph. The Chat Interface enables users to input research questions, retrieve responses generated through GraphRAG, and maintain a conversation history. The Knowledge Graph tab provides a visual representation of retrieved entities and relationships.

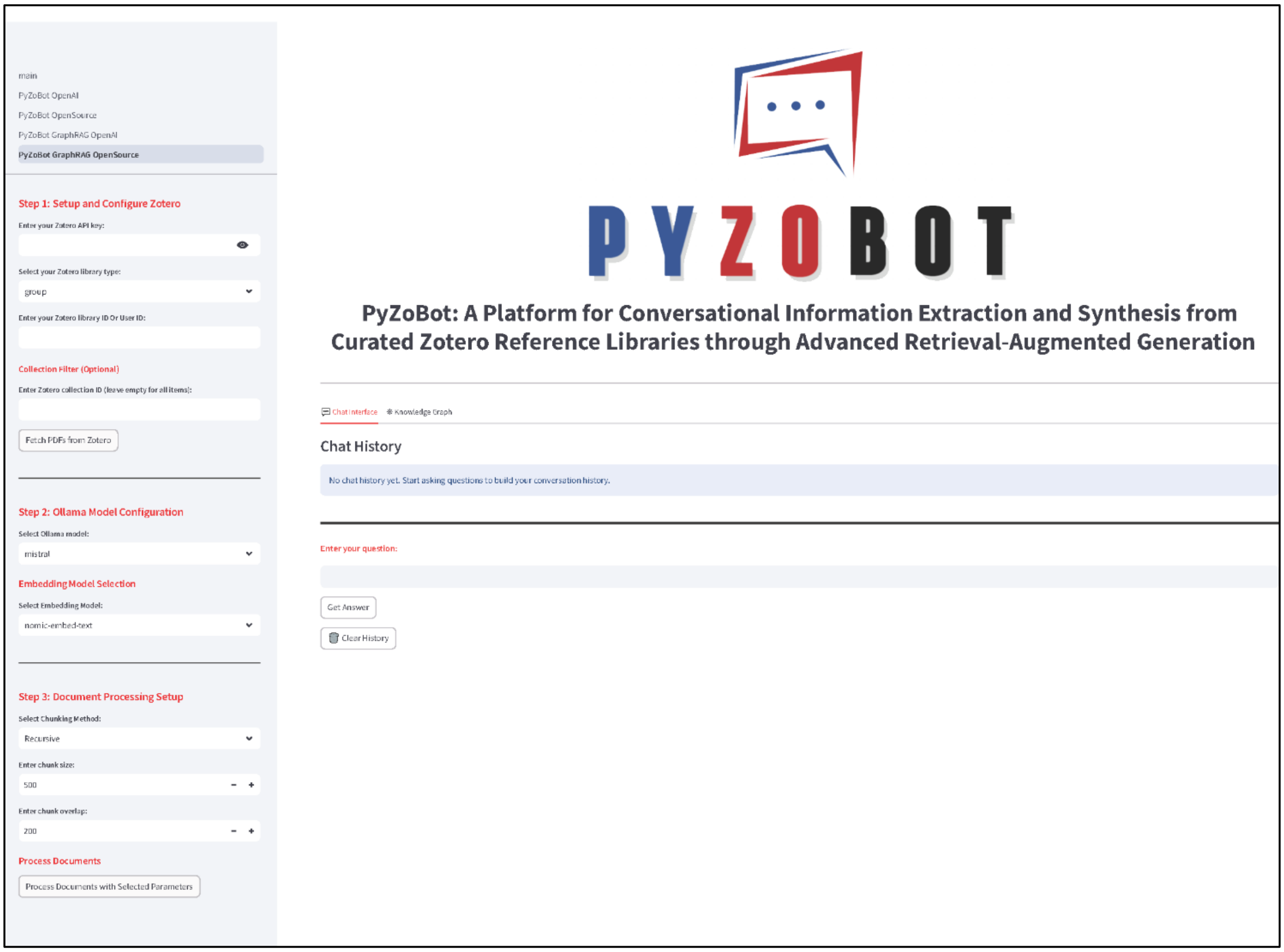

**Figure 8: PyZoBot GraphRAG Open-Source User Interface.** This figure shows the user interface of PyZoBot GraphRAG Open-Source. The left sidebar provides configuration options for Zotero integration, selection of an open-source LLM model (such as LLaMA 3.1 or Mistral), and customization of document processing settings, including chunking methods and retrieval parameters. The query interaction panel is divided into two tabs: Chat Interface and Knowledge Graph. The Chat Interface enables users to input research questions, retrieve responses generated through GraphRAG, and maintain a conversation history. The Knowledge Graph tab provides a visual representation of retrieved entities and relationships.

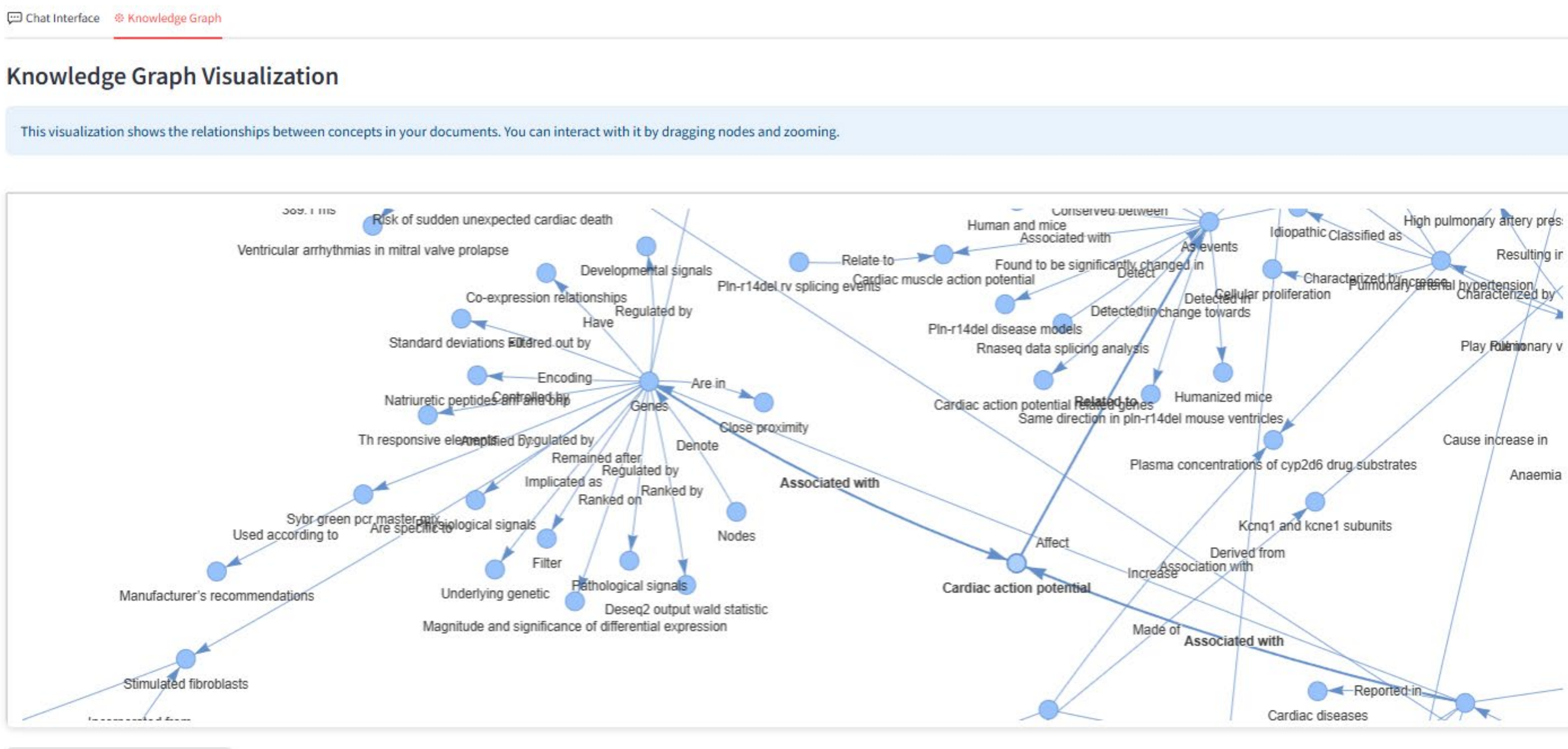

**Figure 9: Knowledge Graph Visualization Constructed by PyZoBot-OpenAI.**

This figure displays a knowledge graph generated using the PyZoBot-OpenAI framework, which integrates GraphRAG techniques with OpenAI language models. The graph was constructed from 22 biomedical research articles sourced from a curated Zotero library (which was used for the yes/no question tasks in Chapter 5). Each node represents an extracted biomedical entity, while edges denote semantic relationships identified between entities. The graph provides an interactive interface that allows users to explore complex biomedical knowledge structures by navigating through concept-to-concept connections. This visualization supports improved retrieval and interpretability of domain-specific information.

An example demonstrating how PyZoBot generates responses through the chat interface and knowledge graph is included in the supplementary files (Supplementary File 1& 2– Conversation Example [PDF] and Supplementary File 3 – Knowledge Graph Example [HTML]). Supplementary File 1 and 2 provides a sample conversation illustrating how PyZoBot answers biomedical queries using retrieved reference PDFs from the Zotero library. Supplementary File 2 displays the corresponding knowledge graph view generated from the retrieved PDFs.

## Conclusion:

The successful implementation of the four PyZoBot variants—two standard RAG systems and two knowledge graph-enhanced RAG systems—demonstrates the feasibility of integrating retrieval-augmented generation techniques with curated Zotero reference libraries. These implementations provide researchers with efficient and customizable tools for information synthesis. The user-friendly interface, modular architecture, and diverse retrieval methods contribute to an adaptable platform that can support a wide range of academic and scientific research needs.

While PyZoBot is designed to streamline literature retrieval and synthesis, a key future direction involves evaluating its performance against that of human researchers conducting similar tasks manually. Such a comparative study will provide valuable insights into the capabilities and limitations of AI-driven information synthesis and help determine the extent to which tools like PyZoBot can effectively support or complement human research workflows.

Code Availability

The complete implementation of PyZoBot, including all four variants (RAG and GraphRAG with both OpenAI and open-source models), is available at https://github.com/dayanjan/PyZoBot-RAG with permanent archival at https://doi.org/10.5281/zenodo.16988871. The repository includes installation instructions, usage examples, and documentation.

## References:

1. Arnold M, Goldschmitt M, Rigotti T. Dealing with information overload: a comprehensive review. *Front Psychol*. 2023;14. doi:10.3389/fpsyg.2023.1122200

2. Eppler M, Mengis J. The Concept of Information Overload: A Review of Literature From Organization Science, Accounting, Marketing, MIS, and Related Disciplines. *Inf Soc*. 2004;20:325-344. doi:10.1080/01972240490507974

3. Bornmann L, Mutz R. Growth rates of modern science: A bibliometric analysis based on the number of publications and cited references: Growth Rates of Modern Science: A Bibliometric Analysis Based on the Number of Publications and Cited References. *Journal of the Association for Information Science and Technology*. 2014;66. doi:10.1002/asi.23329

4. Ghasemi A, Mirmiran P, Kashfi K, Bahadoran Z. Scientific Publishing in Biomedicine: A Brief History of Scientific Journals. *Int J Endocrinol Metab*. 2022;21(1):e131812. doi:10.5812/ijem-131812

5. González-Márquez R, Schmidt L, Schmidt BM, Berens P, Kobak D. The landscape of biomedical research. Published online April 11, 2023:2023.04.10.536208. doi:10.1101/2023.04.10.536208

6. Landhuis E. Scientific literature: Information overload. *Nature*. 2016;535(7612):457-458. doi:10.1038/nj7612-457a

7. Bornmann L, Haunschild R, Mutz R. Growth rates of modern science: A latent piecewise growth curve approach to model publication numbers from established and new literature databases. Published online September 21, 2021. doi:10.48550/arXiv.2012.07675

8. Pautasso M. Ten Simple Rules for Writing a Literature Review. *PLOS Computational Biology*. 2013;9(7):e1003149. doi:10.1371/journal.pcbi.1003149

9. Alshammari S, Basalelah L, Rukbah WA, Alsuhibani A, Wijesinghe DS. KNIMEZoBot: Enhancing Literature Review with Zotero and KNIME OpenAI Integration using Retrieval-Augmented Generation. Published online November 7, 2023. doi:10.48550/arXiv.2311.04310

10. Wagner G, Lukyanenko R, Pare G. Artificial intelligence and the conduct of literature reviews. *Journal of Information Technology*. Published online June 9, 2022:1-18. doi:10.1177/02683962211048201

11. Kousha K, Thelwall M. Artificial intelligence to support publishing and peer review: A summary and review. *Learned Publishing*. 2024;37(1):4-12. doi:10.1002/leap.1570

12. Borgeaud S, Mensch A, Hoffmann J, et al. Improving Language Models by Retrieving from Trillions of Tokens. In: *Proceedings of the 39th International Conference on Machine Learning*. PMLR; 2022:2206-2240. Accessed April 8, 2024. https://proceedings.mlr.press/v162/borgeaud22a.html

13. Thapa S, Adhikari S. ChatGPT, Bard, and Large Language Models for Biomedical Research: Opportunities and Pitfalls. *Ann Biomed Eng*. Published online June 16, 2023. doi:10.1007/s10439-023-03284-0

14. Reese JT, Danis D, Caufield JH, et al. On the limitations of large language models in clinical diagnosis. *medRxiv*. Published online February 26, 2024:2023.07.13.23292613. doi:10.1101/2023.07.13.23292613

15. Alqahtani T, Badreldin HA, Alrashed M, et al. The emergent role of artificial intelligence, natural learning processing, and large language models in higher education and research. *Res Social Adm Pharm*. 2023;19(8):1236-1242. doi:10.1016/j.sapharm.2023.05.016

16. Jungwirth D, Haluza D. Artificial Intelligence and Public Health: An Exploratory Study. *Int J Environ Res Public Health*. 2023;20(5):4541. doi:10.3390/ijerph20054541

17. Pan S, Luo L, Wang Y, Chen C, Wang J, Wu X. Unifying Large Language Models and Knowledge Graphs: A Roadmap. *IEEE Trans Knowl Data Eng*. Published online 2024:1-20. doi:10.1109/TKDE.2024.3352100

18. Eggmann F, Weiger R, Zitzmann NU, Blatz MB. Implications of large language models such as ChatGPT for dental medicine. *J Esthet Restor Dent*. 2023;35(7):1098-1102. doi:10.1111/jerd.13046

19. Ge J, Sun S, Owens J, et al. Development of a Liver Disease-Specific Large Language Model Chat Interface using Retrieval Augmented Generation. *medRxiv*. Published online November 10, 2023:2023.11.10.23298364. doi:10.1101/2023.11.10.23298364

20. Briganti G. How ChatGPT works: a mini review. *Eur Arch Otorhinolaryngol*. 2024;281(3):1565-1569. doi:10.1007/s00405-023-08337-7

21. An H, Narechania A, Wall E, Xu K. vitaLITy 2: Reviewing Academic Literature Using Large Language Models. Published online August 24, 2024. doi:10.48550/arXiv.2408.13450

22. Akinci D'Antonoli T, Stanzione A, Bluethgen C, et al. Large language models in radiology: fundamentals, applications, ethical considerations, risks, and future directions. *Diagn Interv Radiol*. 2024;30(2):80-90. doi:10.4274/dir.2023.232417

23. Chen J, Lin H, Han X, Sun L. Benchmarking Large Language Models in Retrieval-Augmented Generation. Published online 2023. doi:10.48550/ARXIV.2309.01431

24. Gao Y, Xiong Y, Gao X, et al. Retrieval-Augmented Generation for Large Language Models: A Survey. Published online March 27, 2024. doi:10.48550/arXiv.2312.10997

25. Han B, Susnjak T, Mathrani A. Automating Systematic Literature Reviews with Retrieval-Augmented Generation: A Comprehensive Overview. *Applied Sciences*. 2024;14(19):9103. doi:10.3390/app14199103

26. Lorica B. Best Practices in Retrieval Augmented Generation. Gradient Flow. October 19, 2023. Accessed January 29, 2024. https://gradientflow.com/best-practices-in-retrieval-augmented-generation/

27. Retrieval-Augmented Generation & RAG Workflows. Nanonets Intelligent Automation, and Business Process AI Blog. October 25, 2023. Accessed January 29, 2024. https://nanonets.com/blog/retrieval-augmented-generation/

28. Salemi A, Zamani H. Evaluating Retrieval Quality in Retrieval-Augmented Generation. Published online April 21, 2024. doi:10.48550/arXiv.2404.13781

29. Kim T. Building student proficiency with scientific literature using the Zotero reference manager platform. *Biochem Mol Biol Educ*. 2011;39(6):412-415. doi:10.1002/bmb.20551

30. Coar JT, Sewell JP. Zotero: Harnessing the Power of a Personal Bibliographic Manager. *Nurse Educator*. 2010;35(5):205. doi:10.1097/NNE.0b013e3181ed81e4

31. Ahmed KKM, Al Dhubaib BE. Zotero: A bibliographic assistant to researcher. *J Pharmacol Pharmacother*. 2011;2(4):303-305. doi:10.4103/0976-500X.85940

32. Baliyan A, Kaswan KS, Dhatterwal JS. An Empirical Analysis of Python Programming for Advance Computing. In: *2022 2nd International Conference on Advance Computing and Innovative Technologies in Engineering (ICACITE)*. ; 2022:1482-1486. doi:10.1109/ICACITE53722.2022.9823643

33. Stančin I, Jović A. An overview and comparison of free Python libraries for data mining and big data analysis. In: *2019 42nd International Convention on Information and Communication Technology, Electronics and Microelectronics (MIPRO)*. ; 2019:977-982. doi:10.23919/MIPRO.2019.8757088

34. Zhang Y. Comparison of Select Reference Management Tools. *Medical Reference Services Quarterly*. 2012;31(1):45-60. doi:10.1080/02763869.2012.641841

35. Luan A, Momeni A, Lee GK, Galvez MG. Cloud-Based Applications for Organizing and Reviewing Plastic Surgery Content. *Eplasty*. 2015;15:e48.

36. Chen PY, Hayes E, Larivière V, Sugimoto CR. Social reference managers and their users: A survey of demographics and ideologies. *PLoS One*. 2018;13(7):e0198033. doi:10.1371/journal.pone.0198033

37. Liu F, Kang Z, Han X. Optimizing RAG Techniques for Automotive Industry PDF Chatbots: A Case Study with Locally Deployed Ollama Models. Published online August 12, 2024. doi:10.48550/arXiv.2408.05933

38. ollama/ollama. Published online March 4, 2025. Accessed March 4, 2025. https://github.com/ollama/ollama

39. ollama/ollama-python. Published online March 4, 2025. Accessed March 4, 2025. https://github.com/ollama/ollama-python

40. Building a Local RAG System for Privacy Preservation with Ollama and Weaviate | Weaviate. May 21, 2024. Accessed March 4, 2025. https://weaviate.io/blog/local-rag-with-ollama-and-weaviate

41. How to Run Open Source LLMs on Your Own Computer Using Ollama. freeCodeCamp.org. December 20, 2024. Accessed March 4, 2025. https://www.freecodecamp.org/news/how-to-run-open-source-llms-on-your-own-computer-using-ollama/

42. Mishra A. Five Levels of Chunking Strategies in RAG| Notes from Greg's Video. Medium. January 15, 2024. Accessed January 22, 2024. https://medium.com/@anuragmishra_27746/five-levels-of-chunking-strategies-in-rag-notes-from-gregs-video-7b735895694d

43. Recursively split by character | 🦜🔗 Langchain. Accessed January 22, 2024. https://python.langchain.com/docs/modules/data_connection/document_transformers/recursive_text_splitter

44. Understanding LangChain's RecursiveCharacterTextSplitter. DEV Community. August 12, 2023. Accessed March 5, 2025. https://dev.to/eteimz/understanding-langchains-recursivecharactertextsplitter-2846

45. A Guide to Chunking Strategies for Retrieval Augmented Generation (RAG) — Sagacify. Accessed March 5, 2025. https://www.sagacify.com/news/a-guide-to-chunking-strategies-for-retrieval-augmented-generation-rag

46. Kshirsagar A. Enhancing RAG Performance Through Chunking and Text Splitting Techniques. *International Journal of Scientific Research in Computer Science, Engineering and Information Technology*. 2024;10(5):151-158. doi:10.32628/CSEIT2410593

47. Partitioning. Unstructured. Accessed March 5, 2025. https://docs.unstructured.io/open-source/core-functionality/partitioning

48. Partitioning strategies. Unstructured. Accessed March 5, 2025. https://docs.unstructured.io/platform-api/partition-api/partitioning

49. Verma P. S2 Chunking: A Hybrid Framework for Document Segmentation Through Integrated Spatial and Semantic Analysis. Published online January 8, 2025. doi:10.48550/arXiv.2501.05485

50. (PDF) Enhancing RAG Performance Through Chunking and Text Splitting Techniques. *ResearchGate*. Published online November 4, 2024. doi:10.32628/CSEIT2410593

51. Singh IS, Aggarwal R, Allahverdiyev I, et al. ChunkRAG: Novel LLM-Chunk Filtering Method for RAG Systems. Published online November 19, 2024. doi:10.48550/arXiv.2410.19572

52. How to split text based on semantic similarity | 🦜🔗 LangChain. Accessed March 5, 2025. https://python.langchain.com/docs/how_to/semantic-chunker/

53. Sukla A. Using Document Layout Structure for Efficient RAG. Ambika's Substack. October 18, 2023. Accessed March 5, 2025. https://ambikasukla.substack.com/p/efficient-rag-with-document-layout

54. Qu R, Tu R, Bao F. Is Semantic Chunking Worth the Computational Cost? Published online October 16, 2024. doi:10.48550/arXiv.2410.13070

55. Lee J, Cha H, Hwangbo Y, Cheon W. Enhancing Large Language Model Reliability: Minimizing Hallucinations with Dual Retrieval-Augmented Generation Based on the Latest Diabetes Guidelines. *Journal of Personalized Medicine*. 2024;14(12):1131. doi:10.3390/jpm14121131

56. Leng Q, Portes J, Havens S, Zaharia M, Carbin M. Long Context RAG Performance of Large Language Models. Published online November 5, 2024. doi:10.48550/arXiv.2411.03538

57. Nussbaum Z, Morris JX, Duderstadt B, Mulyar A. Nomic Embed: Training a Reproducible Long Context Text Embedder. Published online February 3, 2025. doi:10.48550/arXiv.2402.01613

58. Špeletić M, Špeletić M, Protić M, Nikolić S. EXPLORING RAG IN MEDICAL QUESTION ANSWERING: INTEGRATING LLMS AND VECTOR DATABASES. *UNSP*. Published online 2024. doi:10.70456/GAJM2853

59. BGE Series — BGE documentation. Accessed March 6, 2025. https://bge-model.com/tutorial/1_Embedding/1.2.1.html

60. Chen J, Xiao S, Zhang P, Luo K, Lian D, Liu Z. BGE M3-Embedding: Multi-Lingual, Multi-Functionality, Multi-Granularity Text Embeddings Through Self-Knowledge Distillation. Published online June 28, 2024. doi:10.48550/arXiv.2402.03216

61. Hsain A, Housni HE. Large language model-powered chatbots for internationalizing student support in higher education. Published online March 16, 2024. doi:10.48550/arXiv.2403.14702

62. Albayrak A, Xiao Y, Mukherjee P, Barnett SS, Marcou CA, Hart SN. Enhancing human phenotype ontology term extraction through synthetic case reports and embedding-based retrieval: A novel approach for improved biomedical data annotation. *J Pathol Inform*. 2025;16:100409. doi:10.1016/j.jpi.2024.100409

63. Patel S, Patel J, Shah D, Goel P, Patel B. A RAG based Personal Placement Assistant System using Large Language Models for Customized Interview Preparation. In: *2024 5th International Conference on Data Intelligence and Cognitive Informatics (ICDICI)*. ; 2024:1468-1475. doi:10.1109/ICDICI62993.2024.10810811

64. Improving Document Retrieval with Contextual Compression. LangChain Blog. April 21, 2023. Accessed January 15, 2025. https://blog.langchain.dev/improving-document-retrieval-with-contextual-compression/

65. How to do retrieval with contextual compression | 🦜🔗 LangChain. Accessed January 15, 2025. https://python.langchain.com/docs/how_to/contextual_compression/

66. Alkhalaf M, Yu P, Yin M, Deng C. Applying generative AI with retrieval augmented generation to summarize and extract key clinical information from electronic health records. *Journal of Biomedical Informatics*. 2024;156:104662. doi:10.1016/j.jbi.2024.104662

67. Maryamah M, Irfani MM, Tri Raharjo EB, Rahmi NA, Ghani M, Raharjana IK. Chatbots in Academia: A Retrieval-Augmented Generation Approach for Improved Efficient Information

Access. In: *2024 16th International Conference on Knowledge and Smart Technology (KST)*. ; 2024:259-264. doi:10.1109/KST61284.2024.10499652

68. Gao H, Zhang Y. VRSD: Rethinking Similarity and Diversity for Retrieval in Large Language Models. Published online November 14, 2024. doi:10.48550/arXiv.2407.04573

69. Bora A, Cuayáhuitl H. Systematic Analysis of Retrieval-Augmented Generation-Based LLMs for Medical Chatbot Applications. *Machine Learning and Knowledge Extraction*. 2024;6(4):2355-2374. doi:10.3390/make6040116

70. Kunze KN, Nwachukwu BU, Cote MP, Ramkumar PN. Large Language Models Applied to Health Care Tasks May Improve Clinical Efficiency, Value of Care Rendered, Research, and Medical Education. *Arthroscopy*. 2025;41(3):547-556. doi:10.1016/j.arthro.2024.12.010

71. Cheng J. Applications of Large Language Models in Pathology. *Bioengineering (Basel)*. 2024;11(4):342. doi:10.3390/bioengineering11040342

72. Alber DA, Yang Z, Alyakin A, et al. Medical large language models are vulnerable to data-poisoning attacks. *Nat Med*. Published online January 8, 2025:1-9. doi:10.1038/s41591-024-03445-1

73. Klang E, Tessler I, Apakama DU, et al. Assessing Retrieval-Augmented Large Language Model Performance in Emergency Department ICD-10-CM Coding Compared to Human Coders. *medRxiv*. Published online October 17, 2024:2024.10.15.24315526. doi:10.1101/2024.10.15.24315526

74. Soong D, Sridhar S, Si H, et al. Improving accuracy of GPT-3/4 results on biomedical data using a retrieval-augmented language model. Published online May 30, 2023. doi:10.48550/arXiv.2305.17116

75. Voinea Ștefan V, Mămuleanu M, Teică RV, Florescu LM, Selișteanu D, Gheonea IA. GPT-Driven Radiology Report Generation with Fine-Tuned Llama 3. *Bioengineering (Basel)*. 2024;11(10):1043. doi:10.3390/bioengineering11101043

76. Woo JJ, Yang AJ, Olsen RJ, et al. Custom Large Language Models Improve Accuracy: Comparing Retrieval Augmented Generation and Artificial Intelligence Agents to Noncustom Models for Evidence-Based Medicine. *Arthroscopy*. 2025;41(3):565-573.e6. doi:10.1016/j.arthro.2024.10.042

77. Akyon SH, Akyon FC, Camyar AS, Hızlı F, Sari T, Hızlı Ş. Evaluating the Capabilities of Generative AI Tools in Understanding Medical Papers: Qualitative Study. *JMIR Med Inform*. 2024;12:e59258. doi:10.2196/59258

78. Ivanisenko TV, Demenkov PS, Ivanisenko VA. An Accurate and Efficient Approach to Knowledge Extraction from Scientific Publications Using Structured Ontology Models, Graph Neural Networks, and Large Language Models. *Int J Mol Sci*. 2024;25(21):11811. doi:10.3390/ijms252111811

79. Han H, Wang Y, Shomer H, et al. Retrieval-Augmented Generation with Graphs (GraphRAG). Published online January 8, 2025. doi:10.48550/arXiv.2501.00309

80. Han H, Shomer H, Wang Y, et al. RAG vs. GraphRAG: A Systematic Evaluation and Key Insights. Published online February 17, 2025. doi:10.48550/arXiv.2502.11371

81. Wu J, Zhu J, Qi Y, et al. Medical Graph RAG: Towards Safe Medical Large Language Model via Graph Retrieval-Augmented Generation. Published online October 15, 2024. doi:10.48550/arXiv.2408.04187

82. Property graph - LlamaIndex. Accessed March 10, 2025. https://docs.llamaindex.ai/en/stable/api_reference/indices/property_graph/

83. Knowledge Graph Index - LlamaIndex. Accessed March 6, 2025. https://docs.llamaindex.ai/en/stable/examples/index_structs/knowledge_graph/KnowledgeGraphDemo/

84. PhD KS. Customizing Knowledge Graph Generation and Querying for Improved Response. Medium. July 29, 2024. Accessed March 10, 2025. https://ai.gopubby.com/customizing-knowledge-graph-generation-and-querying-for-improved-response-ef2e269f4395

85. Introducing the Property Graph Index: A Powerful New Way to Build Knowledge Graphs with LLMs — LlamaIndex - Build Knowledge Assistants over your Enterprise Data. Accessed March 6, 2025. https://www.llamaindex.ai/blog/introducing-the-property-graph-index-a-powerful-new-way-to-build-knowledge-graphs-with-llms